\documentclass[prd,a4paper,showpacs,preprint,byrevtex]{revtex4}
\usepackage{graphicx}
\usepackage{dcolumn}
\usepackage{amsmath}
\usepackage{array}
\usepackage{bm}
\usepackage{amssymb}

\begin{document}
%\preprint{Preprint Universit\'{e} de Mons-Hainaut}

\title{Relativistic corrections for two- and three-body flux tube model}

\author{Fabien \surname{Buisseret}}
%\thanks{FNRS Research Fellow}
\email[E-mail: ]{fabien.buisseret@umh.ac.be}
\author{Claude \surname{Semay}}
%\thanks{FNRS Research Associate}
\email[E-mail: ]{claude.semay@umh.ac.be}
\affiliation{Groupe de Physique Nucl\'{e}aire Th\'{e}orique,
Universit\'{e} de Mons-Hainaut,
Acad\'{e}mie universitaire Wallonie-Bruxelles,
Place du Parc 20, BE-7000 Mons, Belgium}

\date{\today}

\begin{abstract}
We generalize the relativistic flux tube model for arbitrary two- or three-body systems. The spin-independent and spin-dependent contributions of the flux tube to the total Hamiltonian are computed in perturbation. In particular, we show that the spin-dependent part exhibits a universal spin-orbit form: It does not depend on the nature of the confined particles. The general equations we present, being well-defined for light particles, can thus be applied to usual as well as exotic hadrons such as hybrid mesons and glueballs. 
\end{abstract}

\pacs{12.39.Pn, 12.39.Ki, 12.39.Mk}
% 12.39.Pn    Potential model
% 12.39.Ki    Relativistic quark model
% 12.39.Mk    Glueball and nonstandard multiquark/gluon states
\keywords{Potential model; Relativistic quark model}

\maketitle
\section{Introduction}

A successful way of understanding the mesons is to approximate the gluon exchanges between the quark and the antiquark by a straight string, which is responsible for the confining interaction. The relativistic flux tube model (RFTM) is an effective QCD model based on this picture \cite{tf_1}. Apart from mesons, it has been generalized to baryons \cite{bramb95}, and to more exotic particles like glueballs and gluelumps (gluon attached to a point-like $q\bar{q}$ pair) \cite{new}. As the present application domain of the RFTM has exceeded its original formulation, it is interesting to explicitly write its equations for general two- or three-body systems, in order to apply it to some cases of current interest like hybrid mesons seen as $q\bar q g$ states, three-gluon glueballs, etc. It is done in Sec.~\ref{sicorr}, where it is assumed that the dynamical contribution of the flux tube is small enough to be treated in perturbation. This approach, that we previously called the perturbative flux tube model (PFTM), is rather satisfactory since it reproduces the exact RFTM spectrum up to $5\%$~\cite{buis05a,bada02}.

The RFTM and the PFTM neglect the spin of the quarks. Recently, an attempt to exactly include spinning particles in the RFTM has been made~\cite{olss04}, but it quickly leads to very complicated equations. In this work, we rather compute these spin contributions within the PFTM framework in order to get more tractable expressions. Similar calculations have already been performed for mesons and baryons with the Wilson loop technique and the background perturbation theory~\cite{barc89,Simo00}. But, the originality of our approach, inspired by Ref.~\cite{Ols}, is that the spin correction we obtain remains valid for particles of arbitrary spin (not only $1/2$ as for the quarks) as well as of arbitrary mass (not only for heavy quarks, but also for light and massless bodies). It has a universal spin-orbit form, which is computed in Sec.~\ref{sdcorr}. Some conclusions are finally drawn in Sec.~\ref{conclu}. 

\section{Perturbative flux tube}\label{sicorr}
\subsection{Two-body systems}

When the orbital angular momentum $\ell$ is equal to zero, the two-body RFTM reduces to the spinless Salpeter Hamiltonian (SSH) $H_0 = \sqrt{\vec p^{\, 2}_1+m_1^2} + \sqrt{\vec p^{\, 2}_2+m_2^2}  + a r$, where the linear confining potential $ar$ is generated by the straight flux tube linking the quark to the antiquark \cite{tf_1}, $a$ being the energy density of the flux tube. For not too large $\ell$, i.e. $\ell<6$, the dynamical contribution of the flux tube can accurately be treated as a perturbation of $H_0$.  The auxiliary field technique allows to compute this perturbation, which is given by \cite{bada02,buis05a,new,ch_aux_bib},
\begin{equation}
\label{hstr1}
\Delta H_{\rm ft} = -\frac{a \ell(\ell+1)}{2\mu_1 \mu_2r} 
\frac{\left[ 4(\mu_1^2 + \mu_2^2 - \mu_1 \mu_2) + (\mu_1 + \mu_2) a
r\right]}{ \left[ 12 \mu_1 \mu_2 + 4 (\mu_1 + \mu_2) a
r + (a r)^2\right]}.
\end{equation}
In this equation, $\mu_i$ can be interpreted as the constituent mass of the particle whose bare mass is $m_{i}$. It reads 
\begin{equation}\label{mu}
\mu_i = \left\langle \sqrt{\vec p^{\, 2}_i+m_i^2}\right\rangle,
\end{equation}
the average being computed with the eigenstates of the unperturbed Hamiltonian $H_0$. Let us notice that $\mu_i>0$ even if $m_i=0$, as for $u,d$ quarks and gluons. It ensures that the correction~(\ref{hstr1}) will be defined in every case. In the heavy quark limit where $\mu_i\approx m_i \gg a r$, the flux tube contribution becomes
\begin{equation}
\label{hstr2}
\Delta H_{\rm ft} \approx -\frac{a \ell(\ell+1)}{6r} 
\left[\frac{1}{\mu_1^2}-\frac{1}{\mu_1 \mu_2}+\frac{1}{\mu_2^2}\right],
\end{equation}
in agreement with the spin-independent correction to potential $ar$ as predicted by the Wilson loop technique~\cite{barc89}. Our formula~(\ref{hstr1}) actually generalizes Eq.~(\ref{hstr2}) to the case of light quarks. It is worth mentioning that, although the discussion we made was based on the mesonic case only, the PFTM Hamiltonian, given by $H=H_0+\Delta H_{\rm ft}$, can be applied to successfully describe glueballs and gluelumps~\cite{new,kaid}. 

\subsection{Three-body systems}\label{3bs}

Lattice QCD calculations support the idea that the so-called Y-junction is the more realistic configuration of the static color field in baryons \cite{Koma}. In this scheme, each quark generates a flux tube, the three flux tubes meeting at the point $Y$ which minimizes the total energy $\sum_i a_i r_i$. $a_i$ and $r_i$ are respectively the energy density and the length of the flux tube starting from particle $i$. 

We assume here that the value of $a_i$ is given by the Casimir scaling hypothesis \cite{scaling}. Under that assumption, the energy density $a_i$ of a flux tube is proportional to its quadratic $SU(3)$ Casimir operator. Then, one should have $a_g=(9/4)\, a_q$.
If the three particles are of the same nature (quark, gluon, \dots), the $a_i$ are all equal, and $Y$ is the Toricelli point, minimizing the total length of the three flux tubes. However, the three bodies can be different in general systems. For example, in a hybrid meson seen as a $q\bar q g$ bound state, one can show that, assuming the Casimir scaling hypothesis, the Y-junction is fixed on the gluon \cite{new,kalash}, with two flux tubes linking the gluon to the quark and to the antiquark. 

The equations defining the three-body RFTM in the center of mass (CM) frame read~\cite{bramb95}
\begin{subequations}\label{tf3c}
\begin{equation}\label{P3b}
\vec 0=\sum^3_{i=1}\vec{p}_i=\sum^3_{i=1}\left[\frac{m_i \vec{v}_i}{\sqrt{1-\vec{v}^{\, 2}_i}}+ \int^1_0 d\theta \frac{a_i r_i\, \vec{v}^{\, t}_i}{\sqrt{1-\vec{v}^{\, t\, 2}_i}}\right],
\end{equation}
\begin{equation}\label{H3b}
H=	\sum^3_{i=1}\left[\frac{m_i}{\sqrt{1-\vec{v}^{\, 2}_i}}+ \int^1_0 d\theta \frac{a_i r_i}{\sqrt{1-\vec{v}^{\, t\, 2}_i}}\right].
\end{equation}
\end{subequations}
The different symbols appearing in these relations have to be clarified. Let $\vec{x}_{Y}$ and $\vec{v}_{Y}$ be the position and the velocity of the Y-junction respectively; similarly, $\vec{x}_i$ and $\vec v_i$ are the position and the velocity of particle $i$. Then, $\vec{r}_i=\vec{x}_i-\vec{x}_Y$, and $a_i$ is the energy density of the flux tube linking $Y$ to that particle. Moreover, $\vec{v}^{\, t}_i=\theta \vec{v}_{i\bot}+(1-\theta) \vec{v}_{Yi\bot}$, where $\vec{v}_{i\bot}$ and $\vec{v}_{Yi\bot}$ are the components of $\vec{v}_{i}$ and $\vec{v}_{Y}$ orthogonal to $\vec{r}_i$. In order to find the PFTM Hamiltonian corresponding to Eqs.~(\ref{tf3c}), one should try to apply the auxiliary field formalism as in the two-body case. However, this procedure is too complex here because we are dealing with a three-body problem. What can be done is to neglect the string contribution in Eq.~(\ref{P3b}). This approximation leads to $\vec v_i=\vec p_i/\mu_i$, where $\mu_i$ is again defined by Eq.~(\ref{mu}). Then, a development of Hamiltonian~(\ref{H3b}) at the order $(\vec v^{\, t}_i)^2$ leads to $H=H_0+\Delta H_{\rm ft}$, where $H_0$ is a three-body SSH with an Y-junction potential,
\begin{equation}
H_0=\sum^3_{i=1}\left[\sqrt{\vec{p}^{\, 2}_i+m^2_i}+a_i r_i	\right],
\end{equation}
and where
\begin{equation}\label{hstr4}
\Delta H_{\rm ft}=	\sum^3_{i=1}\left[-\frac{a_i r_i}{6}\left(\frac{\vec{p}^{\, 2}_{i\bot}}{\mu^2_i}+\vec{v}^{\, 2}_{Yi\bot}+\frac{\vec{p}_{i\bot}\, \vec{v}_{Yi\bot}}{\mu_i}\right)\right]
\end{equation}
is the dynamical contribution of the flux tubes. In the limit of heavy quarks, Eq.~(\ref{hstr4}) reduces to the results of Ref.~\cite{bramb95}.
\par This term can be further simplified in two cases, following the position of the Y-junction. If $\vec{x}_Y\neq\vec{x}_i$, as in baryons and three-gluons glueballs, we can assume in good approximation that the Y-junction is located at the CM. This approximation only overestimates the potential energy of the genuine junction by about $5\%$ in most cases~\cite{Bsb04}. Then, $\vec{v}_{Y}\approx\vec{0}$ as we work in the CM frame. By definition, $|\vec{p}_{i\bot}|=|\vec{L}_i|/r_i$ with $\vec{L}_i=\vec r_i\times \vec p_i$ the orbital angular momentum for the particle $i$, and Eq.~(\ref{hstr4}) becomes 
\begin{equation}
\label{dH3c1}
\Delta H_{\rm str}=	\sum^3_{i=1}\left[-\frac{a_i }{6r_i}\frac{\vec{L}^{\, 2}_{i}}{\mu^2_i}\right],
\end{equation}
where $\vec L_i$ and $r_i$ are now relative to the CM. It is worth mentioning that this last formula is a three-body generalization of Eq.~(\ref{hstr2}): If the sum in Eq.~(\ref{dH3c1}) is performed for a two-body system only, formula~(\ref{hstr2}) is recovered.

If the Y-junction is located on one of the three bodies, as it is the case in a hybrid meson \cite{new}, we can arbitrarily set $\vec{x}_Y=\vec{x}_3$. Then, $\vec r_3=\vec{0}$ and $\vec{v}_Y=\vec{v}_3$. Equation~(\ref{hstr4}) can now be written as 
\begin{equation}\label{dH3c2}
\Delta H_{\rm str}=\sum^2_{i=1}\left[-\frac{a_i r_i}{6}\left(\frac{\vec p_{i\bot}^{\, 2}}{\mu^2_i}+\frac{\vec{p}^{\, 2}_{3i\bot}}{\mu^2_3}+\frac{\vec p_{i\bot}\, \vec p_{3i\bot}}{\mu_i\, \mu_3}\right)\right].	
\end{equation}
This expression is more complicated than (\ref{dH3c1}) since it involves $\vec{p}_{3i\bot}$, that is the component of $\vec{p}_3$ which is orthogonal to $\vec{r}_i=\vec x_i-\vec x_3$. Let us notice that the general equation~(\ref{dH3c2}) agrees with the results of Ref.~\cite{kalash} in the limit of static quarks.

Let us note that, when the quarks are not static, it has been shown in Refs.~\cite{sharov} that the Y-junction becomes unstable at the classical level. Two quarks actually tend to form a diquark linked to the third quark by a single flux tube. Whether such an instability remains or not at the quantum level is still a matter of research. Moreover, the relevance of such an instability for resonances is questionable \cite{sharov}. It is also worth mentioning that the Y-picture can give good description of baryon spectra, even for highly excited states \cite{buis07}. We suggest that the predictions of the Y-junction and of the diquark-quark pictures could be compared in a quantized version of our perturbative flux tube model. We leave such a comparison for future works.

\section{Spin-dependent flux tube contribution}\label{sdcorr}

Let us consider a point-like particle of mass $m$ and charge $q$, evolving in the vector potential $A_\mu$ generated by a fixed source. Its equations of motion are different following its spin, but, in every case, the momentum $p_\mu$ has to be replaced by $\pi_\mu=p_\mu-q A_\mu$ in order to take into account the interaction with the external field. In this section, we will focus on particles whose spin is either $1/2$ or $1$ (quarks or gluons). At the quantum level, the corresponding equations can all be written in a Schr\"odinger-like form $i \partial_t \psi={\cal H}\psi$, where $\psi$ is a ``spinor", whose number of components is different following the spin of the considered particle: $4$ for a Dirac spinor, and $6$ for a spin-$1$ particle. In this last case, one obtains the so-called Duffin-Kemmer-Petiau equation~\cite{dkp}.
\par Once Hamiltonian ${\cal H}$ is known, its nonrelativistic limit can be computed thanks to a Foldy-Wouthuysen transformation \cite{FW}. For a spin-$1/2$ particle, one obtains for the positive-energy part of the Hamiltonian~\cite{FW} 
\begin{equation}\label{fw1}
	{\cal H}=m-q A_0+\frac{\vec{\pi}^{\, 2}}{2m}-\frac{q}{m}\vec{S}\vec{B}+\frac{q}{2 m^2}\vec{S}(\vec{\pi}\times\vec{E})+\frac{q}{8 m^2}\vec{\nabla}\vec{E}.
\end{equation}
The symmetrizations of non commuting operators were not written in order to simplify the notations.  $\vec{S}=\vec\sigma/2$ are the spin-$1/2$ matrices, and the electric and magnetic fields are given by $\vec{E}=-\vec{\nabla}A_0,\ \ \vec{B}=\vec{\nabla}\times\vec{A}$.
The Foldy-Wouthuysen Hamiltonian for a particle of spin $1$ has been computed in Ref.~\cite{Case}. Its positive-energy part is   
\begin{equation}\label{fw2}
	{\cal H}=m-q A_0+\frac{\vec{\pi}^{\, 2}}{2m}-\frac{q}{2m}\vec{S}\vec{B},
\end{equation}
with $(S^j)^{kl}=-i\varepsilon^{jkl}$ the spin-$1$ matrices.  
\par Hamiltonians~(\ref{fw1}) and (\ref{fw2}) were computed in the reference frame of the static source under the implicit assumption that $A_\mu$ was an abelian vector potential. So, they could be seen as part of an effective model of QED, but not of QCD. However, this approach can be applied to QCD by an appropriate choice of $A_\mu$, as it has already been shown in Ref.~\cite{Ols}. We will recall here the main points of this work in order to be self-contained. The question is: Can the PFTM be simulated by a particular form of the vector potential? The straight flux tube of the PFTM actually mimics the configuration of the chromoelectric field as it can be observed in lattice QCD calculations~\cite{Koma}. Consequently, in the rest frame of the flux tube, denoted hereafter as the FT frame, the corresponding gluon field is purely chromoelectric. The Faraday tensor in the FT frame is thus of the form
\begin{equation}\label{fcm}
F^{FT}_{\mu\nu}=(\delta_{\mu0}\delta_{\nu r}-\delta_{\nu0}\delta_{\mu r})\ E^a(r)\ 	\frac{\lambda_a}{2},
\end{equation}
with $r$ the distance between the static source and the test particle. The tensor~(\ref{fcm}) is such that 
\begin{equation}\label{commu}
\left[F^{FT}_{\mu\nu},F^{FT}_{\alpha\beta}\right]=i f^{abc}F^{FT}_{a,\mu\nu}F^{FT}_{b,\alpha\beta}\ \frac{\lambda_c}{2}\propto f^{abc}E_a E_b\, \lambda_c=0.	
\end{equation}
Equation~(\ref{commu}) actually shows that the QCD Faraday tensor behaves as a QED one in this special case. Consequently, we just have to search for the most general vector potential ensuring $\vec{E}^{FT}=-a \vec{r}/r,\ \vec{B}^{FT}=\vec{0}$, which corresponds to the linearly rising potential $A^{FT}_0=ar$ giving the static energy of the flux tube. It is shown in Ref.~\cite{Ols} that this condition is fulfilled if, in the static source's frame, which is also the CM frame, 
\begin{equation}
	A_0=a r \sqrt{1+\left[\frac{\vec{p}}{m+ar}\right]^2}\, ,\quad 
	\vec{A}=\frac{ar}{m+ar}\vec{p},
\end{equation}
where $\vec p$  and $r$ are the momentum of the particle and the flux tube length in the CM frame. In the nonrelativistic limit, $ar\ll m$, $|\vec{p}\ |/m\ll 1$, and thus~\cite{Ols}
\begin{equation}\label{EB}
	A_0\approx a r ,\ \ \vec{A}\approx\frac{ar}{m}\vec{p},\ \ \ \vec{E}=-a\frac{\vec{r}}{r},\ \ \vec{B}=\frac{a \vec{L}}{m r}.	
\end{equation}
\par Formula~ (\ref{EB}) can be injected in the Hamiltonians (\ref{fw1}) and (\ref{fw2}). We set $q=1$ because it can be absorbed in a redefinition of $a$. Finally, the spin-orbit terms we were looking for can be collected. For a spin-$1/2$ particle, 
\begin{eqnarray}\label{soc12}
\Delta {\cal H}_{\rm so}&=&-\frac{1}{m}\vec{S}\vec{B}+\frac{1}{2 m^2}\vec{S}(\vec{\pi}\times\vec{E})=-\frac{a \vec{S}\vec{L}}{2m^2r},
\end{eqnarray}
in agreement with Ref.~\cite{Ols}. For a spin-$1$ particle, 
\begin{eqnarray}\label{soc1}
\Delta {\cal H}_{\rm so}&=&-\frac{1}{2m}\vec{S}\vec{B}=-\frac{a \vec{S}\vec{L}}{2m^2r}.	
\end{eqnarray}
\par We can conclude from this discussion that if one takes into account the spin of the particles, a spin-orbit correction must be added to the PFTM. It has a universal spin-orbit form for spin-$1/2$ and spin-$1$ particles, as it can be seen by inspection of Eqs.~(\ref{soc12}) and (\ref{soc1}), and can be thought as a Thomas precession term in the color magnetic field. This relativistic correction is consequently given in the case of a general two-body system by
\begin{equation}\label{sofin}
	\Delta H_{\rm so}=\sum^2_{i=1}\left[-\frac{a_i \vec{L}_i\vec{S}_i}{2m^2_i r_i}	\right]=-\frac{a}{2r}\left[	\frac{\vec{L}\vec{S}_1}{m^2_1}+	\frac{\vec{L}\vec{S}_2}{m^2_2}\right],
\end{equation}
in agreement with the Wilson loop formalism in the heavy quark case~\cite{barc89}. The last term of this equation is only valid in the CM frame, with $r$ the total flux tube length and $\vec L$ the relative orbital angular momentum. Formula~(\ref{sofin}) is clearly not valid for light particles ($u$, $d$ quarks, and gluons). However, another approach leads to corrections which are defined even in the massless case, that is the background perturbation theory~\cite{Simo00}. Within this formalism, it is shown that the spin-dependent corrections in mesons are given in the CM frame by  
\begin{equation}
\Delta H_{SD}=\left(\frac{a}{r}+\frac{2 }{r}\frac{d\, V_1(r)}{dr} \right)	\left[	\frac{\vec{L}\vec{S}_1}{2\mu^2_1}+	\frac{\vec{L}\vec{S}_2}{2\mu^2_2}\right].
\end{equation}
$V_1(r)$ is a complicated function of the quarks correlators which, at large $r$, becomes $\left.V_1(r)\right|_{r\rightarrow\infty}=-a r$. Then,
\begin{equation}\label{wlcor5}
\Delta H_{SD}\approx-\frac{a}{2r}\left[	\frac{\vec{L}\vec{S}_1}{\mu^2_1}+	\frac{\vec{L}\vec{S}_2}{\mu^2_2}\right].
\end{equation}
Equation~(\ref{wlcor5}) is equal to (\ref{soc12}) up to a substitution of $m_i$ by the dynamical quark masses $\mu_i$~(\ref{mu}), which are always nonzero. In agreement with our results, the correction~(\ref{wlcor5}) also holds for gluons \cite{Simo00}.

For a system in which the flux tubes meet at the CM, like a meson, a baryon or a glueball, the spin-orbit contribution is then be given by 
\begin{equation}\label{sofin2}
	\Delta H_{\rm so}=\sum_{i}\left[-\frac{a_i \vec{L}_i\vec{S}_i}{2\mu^2_i r_i}	\right],
\end{equation}
with $r_i$ the distance between the junction point, located at the CM, and the particle $i$. $\vec L_i$ is the orbital angular momentum of particle $i$, relative to the CM. It is worth noting that our formula~(\ref{sofin2}) agrees with the results of Wilson loop technique~\cite{barc89} and background perturbation theory~\cite{Simo00} although it has been established in a totally different way. This achieves to give us confidence in our result. Let us also remark that the spin-orbit term~(\ref{sofin2}) is expected to be particularly relevant for the glueballs, since the spin interactions are very important in this case, but also because of the Casimir scaling of the energy density of the flux tubes. 

Formula~(\ref{sofin2}) has to be modified when the junction is located on one of the particles, say particle $3$, as for hybrid mesons. The spin-orbit correction is then given by
\begin{equation}
	\Delta H^{hyb}_{\rm so}=\sum_{i=1,2}\left[-\frac{a_i\, \vec{{\mathfrak L}}_i\, \vec{S}_i}{2\mu^2_i r_i}	\right],
\end{equation}
with $\vec r_i=\vec x_i-\vec x_3$ and $\vec{{\mathfrak L}}_i=\vec r_i\times \vec p_i$, in agreement with Refs.~\cite{barc89} for heavy particles. 
\par Other studies reveal that the Foldy-Wouthuysen Hamiltonian for a particle of arbitrary half-integer spin has the same formal structure that the one for a spin-$1/2$ particle \cite{niki}, and that the same conclusion holds for arbitrary integral spin particles \cite{case2}. So, we can assume that the spin-orbit term we found is formally valid for any spin, even if the spin-$1/2$ and $1$ are the most relevant cases for our purpose. 

\section{Conclusion}\label{conclu}

We have explicitly written the equations ruling a general two- or three-body relativistic flux tube model, where the dynamical contribution of the flux tube is seen as a perturbation. The unperturbed Hamiltonian is a spinless Salpeter Hamiltonian with a linear confining potential, while the flux tube perturbation is compatible with spin-independent relativistic corrections arising from other effective approaches~\cite{bramb95,Simo00}. Moreover, we have computed the spin-dependent part of the flux tube contribution. It appears to have a universal spin-orbit form, which does not depend on the spin of the confined particles. 

The perturbative flux tube model we presented is expected to reproduce the exact spectrum of the  relativistic flux tube model with an accuracy of about $5\%$~\cite{buis05a,bada02}. We think that this approach, supplemented by appropriate short-range potentials, is an interesting framework to build effective models describing usual \cite{buis07} as well as exotic hadrons. Indeed, our relativistic corrections are more general than the usual ones because they are also valid for light particles. We leave the computation of hadron mass spectra with such a model for future works.

\acknowledgments
The authors would thank the FNRS (Belgium) for financial support.

\end{document}